# Decision support system for distributed manufacturing based on input–output analysis and economic complexity

| Arnault Pachot | Adélaïde Albouy-Kissi | Benjamin Albouy-Kissi | Frédéric Chausse |
| --- | --- | --- | --- |
| arnault.pachot@etu.uca.fr | adelaide.kissi@uca.fr | benjamin.albouy@uca.fr | frederic.chausse@uca.fr |

Université Clermont-Auvergne, CNRS, SIGMA Clermont, Institut Pascal, F-63000 CLERMONT-FERRAND, FRANCE

The disruption of supplies during the Covid-19 crisis has led to shortages but has also shown the adaptability of some companies, which have succeeded in adapting their production chains quickly to produce goods experiencing shortages: hydroalcoholic gel, masks, and medical gowns. These productive jumps from product A to product B are feasible because of the know-how proximity between the two classes of products. The proximities were computed from the analysis of co-exports and resulted in the construction of the product space.

Based on the product space, as well as the customer–supplier relationships resulting from the input–output matrices, we propose a recommender system for companies. The goal is to promote distributed manufacturing by recommending a list of local suppliers to each company. As there is not always a local supplier for a desired product class, we consider the proximity between products to identify, in the absence of a supplier, a substitute supplier able to adapt its production tools to provide the required product. Our experiments are based on French data, from which we build a graph of synergies illustrating the potential productive links between companies.

Finally, we show that our approach offers new perspectives to determine the level of territories' industrial resilience considering potential productive jumps.

## 1 INTRODUCTION

Many countries are seeking to rebuild their production systems, which have been damaged by years of relocation. The recent crisis has shown the importance of countries and their territories maintaining and developing industrial know-how. At the same time, environmental and societal issues are becoming more prevalent and require short circuits, thus favoring distributed manufacturing models. Public authorities and companies would like to promote local synergies, like customer–supplier relationships, to model resilient production ecosystems.

The availability of open data in economics offers new perspectives on the design of decision support tools. In this study, we formalize a recommender system for companies to guide them in their search for local suppliers. We start by presenting the research work that provides the foundation for this study. Based on this analysis, we detail our original decision support system, which exploits both data from input–output models and productive jumps. We see that it is possible to use economic complexity to determine proximities between classes of products and thus make recommendations for productive jumps that firms can make. These productive jumps represent the capacity of a firm that already produces a type A product to produce a type B product with a slight adaptation of its production tools.

We test our models on the French production system, providing each firm with a list of potential suppliers. Finally, we conclude by reflecting on the perspectives offered by this new approach regarding the relocation of know-how and the development of distributed manufacturing. We discuss the interest in considering productive jumps in an indicator of territorial productive resilience.

## 2 PREVIOUS WORK

### 2.1 Input–output analysis

Input–output analysis uses input–output tables to study the interactions between economic sectors. Leontief (1936) developed a model linking the intermediate production and the final consumption for different industries. The model has been applied to supply chains (Albino et al., 2002, 2011; Doll & Shaffer, 2007; Lee & Yoo, 2016;

Seetharaman et al., 2003; Yu, 2018) and the maritime domain (Huang et al., 2015; Morrissey & O'Donoghue, 2013). Yu (2018) presents a review of use cases in both single-region and multi-region modes.

The structure is presented in Table 1. IO tables describe the intermediate demand between each industrial sector, components of the final demand, import ratios for all the components, value added, tax, subsidies, and gross output per industrial sector.

Table 1: IO table structure (Yu, 2018)

| Business sector | Industry 1 | ⋯ | Industry N | Final demand | Total output |
|---|---|---|---|---|---|
| Industry 1 | $Z_{11}$ | ⋯ | $Z_{1N}$ | $F_1$ | $X_1$ |
| ⋮ | | | | ⋮ | ⋮ |
| Industry N | $Z_{N1}$ | ⋯ | $Z_{NN}$ | $F_N$ | $X_N$ |
| Value added | $V_1$ | ⋯ | $V_N$ | | |
| Imports | $m_1$ | ⋯ | $m_N$ | | |
| Total outlays | $X_1$ | ⋯ | $X_N$ | | |

Let $i$ and $j$ be the two economic sectors, and let $z_{ij}$ be the amount of the flow between $i$ and $j$. Let $X_j$ be the output of sector $j$. The associated technical coefficient $a_{ij}$ is expressed as follows:

$$a_{ij} = \frac{z_{ij}}{X_j} \quad (1)$$

Let $F_i$ be the amount of the final demand of sector $i$. Then, the output of sector $i$, named $X_i$, is determined as follows:

$$X_i = \sum_{j=1}^{N} a_{ij} X_j + F_i \quad (2)$$

We can express the outputs by industry sector as a function of the final demand and industrial technologies with identity matrix $I$:

$$X = (I - A)^{-1} F \quad (3)$$

Tables are published by national or international statistics institutes, like Eurostat or the Bureau of Economic Analysis (US).

## 2.2 Economic complexity

The concept of economic complexity developed by Hausmann and Hidalgo (2010), Hidalgo and Hausmann (2006, 2009), and Hidalgo et al. (2007) is based on the analysis of trade between countries. By observing the exports of a country, we can measure its productive know-how. This productive know-how is the key to a country's economic prosperity and may require a complex network of specific skills acquired over several years. A country's economic complexity indicator (ECI) is an indicator of the complexity of its productive know-how. The underlying idea behind the concept of economic complexity is that the evolution of a country's productive know-how takes place progressively by "jumping" from the manufacture of one type of product to another type of product that is close to it in the product space.

Numerous applications of the concept of economic complexity are dedicated to global analysis (Hidalgo et al., 2007), sub-national analysis (Balsalobre & Verduras, 2017; Balsalobre et al., 2019; Reynolds et al., 2018), or societal analysis (Ben Saâd & Assoumou-Ella, 2019; Ferraz et al., 2018; Hartmann et al., 2017; Lapatinas, 2016; Le Caous & Huarng, 2020; Lee & Vu, 2020; Morais et al., 2018; Sbardella et al., 2017; Zhu et al., 2017)0. Other works analyze growth (02014), sustainability (Fraccascia et al., 2018; Hamwey et al., 2013; Huberty & Zachmann, 2011; Mealy & Teytelboym, 2020; Perruchas et al., 2020), diversification capacities (Alshamsi et al., 2018; Boschma & Capone, 2015; Pachot et al., 2021a, 2021b; Zhu et al., 2017), or the impact of public policies (Marrocu et al., 2020; Uhlbach et al., 2017).

Based on the large-scale analysis of the types of products exported by countries, researchers are able to measure the proximity of productive know-how between each type of product and construct a graph of products named the product space (Hidalgo & Hausmann, 2009). The results obtained for each country can be consulted in the Atlas of Economic Complexity.[1]

---
[1] https://atlas.cid.harvard.edu

Let $X_{cp}$ be the exports of product $p$ by country $c$; then, the revealed competitive advantage that country $c$ has for product $p$ can be expressed as a function of exports according to the formula of Belassa (1986):

$$RCA_{cp} = \left(\frac{X_{cp}}{\sum_c X_{cp}}\right) / \left(\frac{\sum_p X_{cp}}{\sum_{c,p} X_{cp}}\right) \quad (4)$$

In the following formulas, country $c$ is considered to export product $p$ if and only if $RCA_{cp}$ is greater than 1. For example, in 2019, wine accounted for 0.136% of world trade (33.8 billion), with total world exports of $24,795 billion. Of this total, Georgia exported nearly $193 million worth of wine. Georgia's total exports for that year amounted to $7.81 billion, of which wine accounted for 2.47%. Georgian wine thus represents 0.57% of wine exports in the world, while Georgian exports of all products combined represent 0.03% of the world exports. We obtain a revealed competitive advantage (RCA) of 18.12 for wine in Georgia, which means that Georgia exports 18.12 times its fair share of wine exports, so we can say that Georgia has a significant revealed competitive advantage for wine.

The calculation of the productive proximity between products is undertaken by looking for each pair of products $\{p_1; p_2\}$ exported together:

$$M_{cp} = \begin{cases} 1 & if \ RCA_{cp} \geq 1 \\ 0 & else \end{cases} \quad (5)$$

$$\phi_{p_1,p_2} = min\left(\frac{\sum_c M_{cp_1} M_{cp_2}}{\sum_c M_{cp_1}}, \frac{\sum_c M_{cp_1} M_{cp_2}}{\sum_c M_{cp_2}}\right) \quad (6)$$

## 3 MODELIZATION

Our decision support system is organized around three processes: the identification of facilities in a territory, the search for potential suppliers for them, and finally the search for alternative suppliers.

### 3.1 Identification of facilities in a territory from open data

From the open data on facilities available in many countries, we can perform mapping work to represent each facility on its territory. Based on its postal address, we can then use geolocation services to obtain the GPS coordinates of the facility.

It is possible to download the data sets directly from the competent organizations in the different countries or to use a platform such as opencorporates.com,[2] which aggregates open data relating to more than 200,000 companies and 276,000 facilities located in a hundred different countries. The facilities are associated with an industry code linked to their main activity. Depending on the country, conversions of economic activity nomenclature are necessary. The Ramon and Unstat[3] websites offer numerous tables of correspondence between nomenclatures.

### 3.2 Looking for potential suppliers for each facility based on input–output matrices

For each facility in a territory, we wish to identify the presence of potential suppliers. Potential suppliers are facilities of which the production enters the intermediate consumption. An interesting approach to identify customer–supplier relationships is to use input–output tables.

*3.2.1 BEA–NACE conversion*

The matrix of dependencies between activities is expressed in the BEA (Bureau of Economic Analysis) nomenclature. This nomenclature is adapted from the NAICS nomenclature of economic activities in North America. Despite this, the correspondence with the latter is not simple because the relationships are of type $1 \rightarrow n$, $n \rightarrow 1$, or $n \rightarrow n$. Several NAICS codes can correspond to one and the same BEA code, and the proportions of breakdowns into these different NAICS codes are not known. It is impossible to use a correspondence table directly without pre-processing because there are no direct transition tables between the product codes used for the composition of the IO tables and the different European (NACE) or international (ISIC) nomenclatures.

---

[2] https://opencorporates.com
[3] https://unstats.un.org/unsd/classifications/Econ

We need a correspondence table between the BEA code and the different NACE classes. We proceed manually by evaluating the correspondences between each class arising from the two nomenclatures, using the information contained in the transition between NACE and NAICS. For each BEA code, we analyze the correspondence with the NAICS classes, then we determine how many NACE classes are associated with each NAICS class. We treat the three cases that can occur separately:

i. If only one match exists between the BEA code and the NAICS code and only one match exists between that same NAICS code and the NACE classification, the transition matrix will have a value of 1 between the BEA class and the designated NACE class.

ii. If only one match exists between the BEA code (and, respectively, the NAICS code) and the NAICS code (and, respectively, the NACE code), but several matches exist between the NAICS (BEA) code and the NACE (NAICS) code, we need to break down the number of occurrences between the NAICS (BEA) code and the NACE (NAICS) code present in the transition between the two nomenclatures. For example, there is only one match between BEA code 111200 "Vegetable and melon farming" and NAICS code 1112 "Vegetable and melon farming" with four digits. At the detailed level, however, the NAICS code 1112 is broken down into four six-digit codes, which are themselves divided into three distinct NACE classes. Here, to move from the BEA code to the NACE code, the breakdown is performed in a linear way. We take the number of occurrences for each NACE code to proceed with the breakdown. In the case of this sector, the weight of each NACE code in the transition matrix uses the following calculation: if $w_i$ is the weight of the sector of activity $i$ in the transition matrix, if $n$ is the number of breakdowns existing between the NAICS code and the different NACE codes, and if $o_i$ is the number of occurrences of the ith NACE code in the breakdown of the NAICS code, then we obtain $w_i = \frac{o_i}{n}$.

iii. The third case arises when the BEA code is broken down into several NAICS codes, which are broken down into several NACE codes. Here, the calculation is as follows:

$$w_i = \frac{1}{n_{NAICS}} \cdot \frac{1}{n_{NACE}} \cdot o_i \quad (7)$$

The only difference here is that the occurrence of industry $i$ is weighted by the number of NAICS codes present in the BEA code breakdown multiplied by the number of NACE codes present in the NAICS code breakdown.

We then obtain a BEA to NACE correspondence table, which allows us to modify the IO matrices to obtain a NACE–NACE association table, which associates $n$ NACE codes in the columns with each NACE code in the rows, corresponding to potential suppliers. We can then easily list, for a given facility for which the NACE code is known, the facilities in the territory that could potentially supply it. This table is available online in the project repository.[4]

### 3.3 Identify alternative suppliers based on productive proximity

To complete the list of suppliers resulting from the previous process with proposals of closer suppliers, or simply to compensate for the absence of potential suppliers, we propose a method to identify alternative suppliers. Alternative suppliers are suppliers that do not currently produce the necessary goods or raw materials but that could, in the short or medium term, adapt their production tools to produce them.

Let $\widehat{R}_{e,\tau}$ be the list of recommended suppliers for facility $e$ in territory $\tau$. Let $\widehat{R}'_{e,\tau}$ be the list of alternative suppliers for the facility in the territory. The final list of recommended suppliers is defined as follows:

$$\widehat{RS}_{e,\tau} = \widehat{R}_{e,\tau} \cup \widehat{R}'_{e,\tau} \quad (8)$$

To identify alternative suppliers, we consider the similarities in know-how between the different product classes. We have seen, during different crises, that companies adapt to produce goods that are in short supply, such as masks, artificial respirators, hydroalcoholic gel, or protective gowns. To measure this level of proximity between two classes of products, we rely on work related to the study of economic complexity.

---

[4] https://github.com/apachot/Decision-support-system-for-distributed-manufacturing

From the website of the Atlas of Economic Complexity, we can download a table of product proximities in HS1992 in four digits.[5] As we intend to work on HS2017, we identify the correspondence between the two nomenclatures.[6]

*3.3.1 Building weighted correspondence between products and economic activity*

We have just studied the notion of a productive relationship between two classes of products; we must now determine the correspondence with the economic activities to measure the proximities between them according to the proximity of the products that they produce. From the industry code of a facility, we need to know the typology of its production, that is, which goods are produced and in what proportion. We can identify the products associated with an industry code by using correspondence tables. To solve the problem of the weighting of each product, we use a statistical approach that analyzes the total amount of exports in euros of each class of products in a territory and deduces the weights. There are several international or national nomenclatures, and we choose the NACE Rev. 2 nomenclature as the reference nomenclature. Similarly, there are several product nomenclatures, and we use the Harmonized System (HS) since this nomenclature has been used previously.

To determine the weights of each product class, we use the amounts of exports. By resorting to a combination of correspondence tables between activities and products (NACE→CPA→HS),[7] we calculate a weight for every product class that is related to an activity, basing our calculation on the export amounts per product. Let $p_1, \dots, p_n$ be the list of products associated with economic activity $a$. Let $X_{cp_i}$ be the total amount of exports of product $p_i$ in country $c$. Then, weight $\lambda_{cap_i}$ of product $p_i$ within activity $a$ in country $c$ is defined as follows:

$$\lambda_{cap_i} = \frac{X_{cp_i}}{\sum_{p_j \in a} X_{cp_j}} \qquad (9)$$

Thus, the output associated with activity $a$ in country $c$ is equivalent to:

$$\lambda_{cap_1} p_1 + \lambda_{cap_2} p_2 + \dots + \lambda_{cap_n} p_n \quad / \quad \sum_{i=1}^{n} \lambda_{cap_i} = 1 \quad (10)$$

*3.3.2 Multidimensional scaling and the measurement of proximities between activities*

We use multidimensional scaling (MDS) to represent the products in a vector space according to their proximities. In our case, we perform a metric MDS, taking as input a symmetric matrix of dissimilarities between products. We construct the dissimilarity matrix from the productive proximities.[8] Let $X_1, \dots, X_n$ be the products and $\delta_{i,j}$ be the distance between product $i$ and product $j$ in the training set. Our stress majorization cost function is:

$$\sigma(X) = \sqrt{\sum_{i=1}^{n} \sum_{j=1, i \neq j}^{n} (\|x_i - x_j\| - \delta_{i,j})^2} \qquad (11)$$

We obtain an *m*-dimensional vector space $E$ in which each product $p_i$ is associated with a vector $v_{p_i}$. From the weighted correspondence table presented previously, we represent the economic activities in the form of vectors, the construction of which is carried out using the weighted average of the vectors of the products that are associated with them. Thus, if products $p_1, \dots, p_n$ are associated with economic activity $a$, and if the weights of each product are represented by variables $\lambda_{cap_1}, \dots, \lambda_{cap_n}$, then vector $v_a$, representing activity $a$, can be written as:

$$\sum_{i=1}^{n} \lambda_{cap_i} v_{p_i} \qquad (12)$$

---

[5] https://dataverse.harvard.edu/dataset.xhtml?persistentId=doi:10.7910/DVN/FCDZBN
[6] https://unstats.un.org/unsd/trade/classifications/tables/HS2017toHS1992ConversionAndCorrelationTables.xlsx
[7] We parse the NAF to the CPF correspondence document (https://www.insee.fr/fr/statistiques/fichier/2399243/Nomenclatures_NAF_et_CPF_Reedition_2020.pdf) to obtain a NACE 2 to CPF 2.1 CSV file. The NACE is obtained from the NAF code by removing the last letter. CPF and CPA are identical in version 2.1. The matching between CPA 2.1 and HS 2017 is performed using the CPA 2.1 to NC 2017 correspondence table (https://ec.europa.eu/eurostat/ramon/relations/index.cfm?TargetUrl=LST_REL) because the HS code is equivalent to the sixth first digits of the related NC code. See 0 for details.
[8] https://dataverse.harvard.edu/dataset.xhtml?persistentId=doi:10.7910/DVN/FCDZBN

This representation opens up the possibility of measuring distances between economic activities in vector space $E$. The proximity between two economic activities $a$ and $b$ is then realized by measuring the inverse of the cosine similarity between their two respective vectors in $E$:

$$\phi_{a,b} = 1 / \frac{v_a \cdot v_b}{|v_a||v_b|} \quad (13)$$

We build a proximity table between each NACE code, which is available from the project repository.[9]

## 4 EXPERIMENTS AND RESULTS

The experiments were carried out on the Atlas of Productive Synergies.[10] We conducted our experiments in France. The data related to companies are available in open data from the Sirene database.[11] For each active facility, we have information about its activity code,[12] its postal address, and its French department. We use the geolocation service to represent the facilities geographically.[13] When the system fails to find the complete address, we progressively simplify the address by deleting some elements until we keep only the name of the street and/or the city and thus obtain the approximate coordinates.

### 4.1 IO matrices for France

These data are available only at the aggregate level for French production. Indeed, only one input–output table is made available by Eurostat.[14] This table is only aggregated into 64 + 1 sectors of activity.

After compiling the French input–output matrix from these data, there are a couple of limitations to its use:

i. Some sectors, particularly industrial sectors such as NACE 20 "Manufacture of chemicals and chemicals products," are far too aggregated and encompass far too many different activities to allow a realistic recommendation of these suppliers.

ii. This database makes it possible to recover information on the construction of the value-added chain not for the production of a product but for an entire sector of activity. This bias does not allow this database to be linked to others collected in the framework of the atlas of productive synergies.

Given these two limitations, we decided to use the U.S. table presented earlier. Obviously, the use of this table requires the assumption that the different value-added chains of products in France are similar to those in the United States. Although this assumption does not accurately reflect the economic reality, we assume this approximation because there are no significant differences in the composition and rank of importance of products used as intermediate consumption in the production of a given good between France and the United States.

### 4.2 Identification of potential suppliers for each French facility

From the IO matrices and geolocation information, we can identify potential suppliers located within a defined perimeter. For each French facility, we build a list of nearby facilities able to produce the manufactured goods or raw materials that are part of its intermediate consumption.

---

[9] https://github.com/apachot/Decision-support-system-for-distributed-manufacturing
[10] https://atlas.productive-synergies.com
[11] https://www.sirene.fr/sirene/public/accueil
[12] The activity code is provided by the French NAF nomenclature, the first four digits of which correspond to the European NACE nomenclature.
[13] https://api-adresse.data.gouv.fr
[14] https://ec.europa.eu/eurostat/web/esa-supply-use-input-tables/data/database

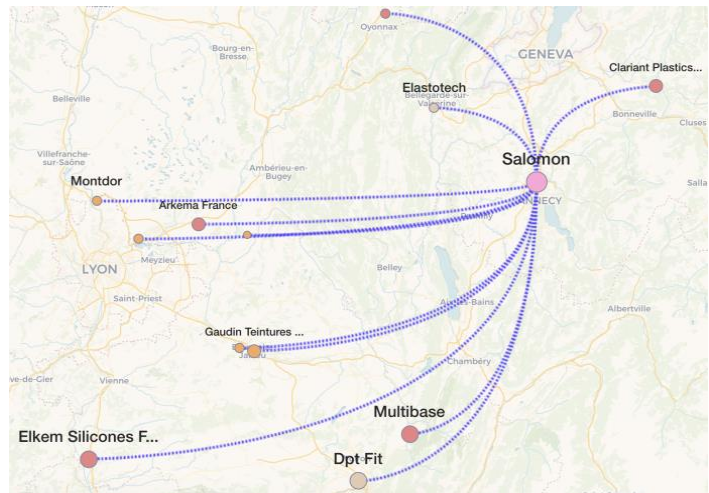

*Figure 1: Example of potential suppliers for the Salomon company (manufacture of sports equipment in Annecy, France)*

In the same way, we compile a second list of alternative suppliers by extending the search to close industry codes, according to the table constructed in paragraph 3.3.2. For each facility, we therefore have two lists of potential suppliers. We obtain a network representing the potential productive synergies between French facilities.

## 5 CONCLUSION AND PERSPECTIVES

In this study, we exploited open data to build a network of potential suppliers associated with a territory. Our main contribution lies in the joint use of input–output matrices and know-how proximity data. We defined a new similarity matrix between two industry codes based on the proximity of their respective vectors in a multidimensional space.

Our work offers a solution to the search for local suppliers for firms and thus promotes distributed manufacturing and the reconstruction of sustainable productive systems. In addition, it opens up new perspectives in the definition of a territorial industrial resilience indicator. The main approaches based on network analysis are focused on the actual production (Craighead et al., 2007; Kharrazi et al., 2017; Kim et al., 2015; Zio & Piccinelli, 2010). However, alternative production is activated in the event of a destabilizing shock, as happened during the last health crisis. The underlying idea is that, in terms of resilience, it may not be necessary to produce certain goods if local suppliers are able to modify their production line quickly to produce them. Suppliers adapt their production to compensate for shortages of certain products, and the integration of this new dimension is interesting to investigate.

## REFERENCES

Albino, V., Izzo, C., & Kühtz, S. (2002). Input–output models for the analysis of a local/global supply chain. *International Journal of Production Economics*, *78*(2), 119–131. https://doi.org/10.1016/S0925-5273(01)00216-X

Albino, V., Petruzzelli, A. M., Okogbaa, G., & Yazan, D. (2011). Logistics flows and enterprise input–output models: Aggregate and disaggregate analysis. *International Journal of Operations and Quantitative Management*, *17*, 123–146.

Alshamsi, A., Pinheiro, F. L., & Hidalgo, C. A. (2018). Optimal diversification strategies in the networks of related products and of related research areas. *Nature Communications*, *9*(1), 1328. https://doi.org/10.1038/s41467-018-03740-9

Balassa, B. (1986). Comparative advantage in manufactured goods: A reappraisal. *Review of Economics and Statistics*, *68*(2), 315. https://doi.org/10.2307/1925512

Balsalobre, S. J. P., & Verduras, C. L. (2017). Measuring the economic complexity at the sub-national level using international and interregional trade. In *Nineteenth Annual Conference of European Trade Study Group* (ETSG 2017), Florence, Italy.

Balsalobre, S. P., Verduras, C. L., & Diaz-Lanchas, J. (2019). *Measuring subnational economic complexity: An application with Spanish data* (JRC Working Papers on Territorial Modelling and Analysis No. 2019-05). Seville: Joint Research Centre. https://ideas.repec.org/p/ipt/termod/201905.html

Ben Saâd, M., & Assoumou-Ella, G. (2019). Economic complexity and gender inequality in education: An empirical study. *SSRN Electronic Journal*, *39*(1), 321–334. https://doi.org/10.2139/ssrn.3340913

Boschma, R., & Capone, G. (2015). Institutions and diversification: Related versus unrelated diversification in a varieties of capitalism framework. *Research Policy*, *44*(10), 1902–1914. https://doi.org/10.1016/j.respol.2015.06.013

Craighead, C. W., Blackhurst, J., Rungtusanatham, M. J., & Handfield, R. B. (2007). The severity of supply chain disruptions: Design characteristics and mitigation capabilities. *Decision Sciences*, *38*(1), 131–156. https://doi.org/10.1111/j.1540-5915.2007.00151.x

Delgado, M., Porter, M. E., & Stern, S. (2014). Clusters, convergence, and economic performance. *Research Policy*, *43*(10), 1785–1799. https://doi.org/10.1016/j.respol.2014.05.007

Doll, C., & Schaffer, A. (2007). Economic impact of the introduction of the German HGV toll system. *Transport Policy*, *14*(1), 49–58.

https://doi.org/10.1016/j.tranpol.2006.09.001

Ferraz, D., Moralles, H. F., Campoli, J. S., Oliveira, F. C. R. de, & Rebelatto, D. A. do N. (2018). Economic complexity and human development: DEA performance measurement in Asia and Latin America. *Gestão & Produção*, *25*(4), 839–853. https://doi.org/10.1590/0104-530x3925-18

Fraccascia, L., Giannoccaro, I., & Albino, V. (2018). Green product development: What does the country product space imply? *Journal of Cleaner Production*, *170*, 1076–1088. https://doi.org/10.1016/j.jclepro.2017.09.190

Hamwey, R., Pacini, H., & Assunção, L. (2013). Mapping green product spaces of nations. *Journal of Environment & Development*, *22*(2), 155–168. https://doi.org/10.1177/1070496513482837

Hartmann, D., Guevara, M. R., Jara-Figueroa, C., Aristarán, M., & Hidalgo, C. A. (2017). Linking economic complexity, institutions, and income inequality. *World Development*, *93*, 75–93. https://doi.org/10.1016/j.worlddev.2016.12.020

Hausmann, R., & Hidalgo, C. A. (2010). *Country diversification, product ubiquity, and economic divergence* (Faculty Research Working Paper Series 10-045). John F. Kennedy School of Government, Harvard University.

Hausmann, R., & Klinger, B. (2006). *Structural transformation and patterns of comparative advantage in the product space* (CID Faculty Working Paper No. 128). Center for International Development, Harvard University.

Hidalgo, C. A., & Hausmann, R. (2009). The building blocks of economic complexity. *Proceedings of the National Academy of Sciences*, *106*(26), 10570–10575. https://doi.org/10.1073/pnas.0900943106

Hidalgo, C. A., Klinger, B., Barabasi, A.-L., & Hausmann, R. (2007). The product space conditions the development of nations. *Science*, *317*(5837), 482–487. https://doi.org/10.1126/science.1144581

Huang, W., Corbett, J. J., & Jin, D. (2015). Regional economic and environmental analysis as a decision support for marine spatial planning in Xiamen. *Marine Policy*, *51*, 555–562. https://doi.org/10.1016/j.marpol.2014.09.006

Huberty, M., & Zachmann, G. (2011, May). *Green exports and the global product space: Prospects for EU industrial policy* (Bruegel Working Paper 2011/07). Breugel.

Kharrazi, A., Rovenskaya, E., & Fath, B. D. (2017). Network structure impacts global commodity trade growth and resilience. *PLOS ONE*, *12*(2), e0171184. https://doi.org/10.1371/journal.pone.0171184

Kim, Y., Chen, Y.-S., & Linderman, K. (2015). Supply network disruption and resilience: A network structural perspective. *Journal of Operations Management*, *33–34*(1), 43–59. https://doi.org/10.1016/j.jom.2014.10.006

Lapatinas, A. (2016). Economic complexity and human development: A note. *Economics Bulletin*, *36*(3), 1441–1452.

Le Caous, E., & Huarng, F. (2020). Economic complexity and the mediating effects of income inequality: Reaching sustainable development in developing countries. *Sustainability*, *12*(5), 2089. https://doi.org/10.3390/su12052089

Lee, K.-K., & Vu, T. V. (2020). Economic complexity, human capital and income inequality: A cross-country analysis. *Japanese Economic Review*, *71*(4), 695–718. https://doi.org/10.1007/s42973-019-00026-7

Lee, M.-K., & Yoo, S.-H. (2016). The role of transportation sectors in the Korean national economy: An input–output analysis. *Transportation Research Part A: Policy and Practice*, *93*, 13–22. https://doi.org/10.1016/j.tra.2016.08.016

Leontief, W. W. (1936). Quantitative input and output relations in the economic systems of the United States. *Review of Economics and Statistics*, *18*(3), 105. https://doi.org/10.2307/1927837

Marrocu, E., Paci, R., Rigby, D., & Usai, S. (2020). *Smart specialization strategy: Any relatedness between theory and practice?* (Working Paper CRENoS No. 202004). Centre for North South Economic Research, University of Cagliari and Sassari. https://ideas.repec.org/p/cns/cnscwp/202004.html

Mealy, P., & Teytelboym, A. (2020). Economic complexity and the green economy. *Research Policy*, 103948. https://doi.org/10.1016/j.respol.2020.103948

Morais, M. B., Swart, J., & Jordaan, J. A. (2018). *Economic complexity and inequality: Does productive structure affect regional wage differentials in Brazil?* (Working Paper No. 18-11). Utrecht School of Economics. https://ideas.repec.org/p/use/tkiwps/1811.html

Morrissey, K., & O'Donoghue, C. (2013). The potential for an Irish maritime transportation cluster: An input–output analysis. *Ocean & Coastal Management*, *71*, 305–313. https://doi.org/10.1016/j.ocecoaman.2012.11.001

Pachot, A., Albouy-Kissi, A., Albouy-Kissi, B., & Chausse, F. (2021a, May). Production2Vec: A hybrid recommender system combining semantic and product complexity approach to improve industrial resiliency. In *2nd International Conference on Artificial Intelligence and Information Systems* (ICAIIS '21). https://doi.org/10.1145/3469213.3469218

Pachot, A., Albouy-Kissi, A., Albouy-Kissi, B., & Chausse, F. (2021b). Multiobjective recommendation for sustainable production systems. In H. Abdollahpouri, M. Elahi, M. Mansoury, S. Sahebi, Z. Nazari, A. Chaney, & B. Loni (Eds.), *Proceedings of the 1st Workshop on Multi-Objective Recommender Systems (MORS 2021) co-located with the 15th ACM Conference on Recommender Systems (RecSys 2021)*, Amsterdam, the Netherlands, September 25, 2021 (Vol. 2959). CEUR-WS.org. http://ceur-ws.org/Vol-2959/paper7.pdf

Perruchas, F., Consoli, D., & Barbieri, N. (2020). Specialisation, diversification and the ladder of green technology development. *Research Policy*, *49*(3), 103922. https://doi.org/10.1016/j.respol.2020.103922

Reynolds, C., Agrawal, M., Lee, I., Zhan, C., Li, J., Taylor, P., Mares, T., Morison, J., Angelakis, N., & Roos, G. (2018). A sub-national economic complexity analysis of Australia's states and territories. *Regional Studies*, *52*(5), 715–726. https://doi.org/10.1080/00343404.2017.1283012

Sbardella, A., Pugliese, E., & Pietronero, L. (2017). Economic development and wage inequality: A complex system analysis. *PLOS ONE*, *12*(9), e0182774. https://doi.org/10.1371/journal.pone.0182774

Seetharaman, A., Kawamura, K., & Dev Bhatta, S. (2003). Economic benefits of freight policy relating to trucking industry: Evaluation of regional transportation plan freight policy for a six-county region, Chicago, Illinois. *Transportation Research Record: Journal of the Transportation Research Board*, *1833*(1), 17–23. https://doi.org/10.3141/1833-03

Uhlbach, W.-H., Balland, P. A., & Scherngell, T. (2017). R&D policy and technological trajectories of regions: Evidence from the EU Framework Programmes. *SSRN Electronic Journal*. https://doi.org/10.2139/ssrn.3027919

Yu, H. (2018). A review of input–output models on multisectoral modelling of transportation–economic linkages. *Transport Reviews*, *38*(5), 654–677. https://doi.org/10.1080/01441647.2017.1406557

Zhu, S., He, C., & Zhou, Y. (2017). How to jump further and catch up? Path-breaking in an uneven industry space. *Journal of Economic Geography*, lbw047. https://doi.org/10.1093/jeg/lbw047

Zio, E., & Piccinelli, R. (2010). Randomized flow model and centrality measure for electrical power transmission network analysis. *Reliability Engineering & System Safety*, *95*(4), 379–385. https://doi.org/10.1016/j.ress.2009.11.008